\documentclass[runningheads]{llncs}
\usepackage{subfig}
\usepackage[T1]{fontenc}
\usepackage{graphicx}
\begin{document}
\title{PatternPortrait: Draw Me Like One of Your Scribbles}

\author{Sabine Wieluch\orcidID{0009-0002-2007-4747} \and \\
	Friedhelm Schwenker\orcidID{0000-0001-5118-0812}}

\authorrunning{S. Wieluch et al.}

\institute{Ulm University, 89069 Ulm, Germany\\
	\email{sabine.wieluch@uni-ulm.de}\\
	\url{https://www.uni-ulm.de/in/neuroinformatik/}}

\maketitle              
\begin{abstract}
This paper introduces a process for generating abstract portrait drawings from pictures. Their unique style is created by utilizing single freehand pattern sketches as references to generate unique patterns for shading. The method involves extracting facial and body features from images and transforming them into vector lines. A key aspect of the research is the development of a graph neural network architecture designed to learn sketch stroke representations in vector form, enabling the generation of diverse stroke variations. The combination of these two approaches creates joyful abstract drawings that are realized via a pen plotter. The presented process garnered positive feedback from an audience of approximately 280 participants.

\end{abstract}
\section{Introduction and Related Work}
What exactly is a portrait? The most common association is a historical drawn painting of a person. Though portraits can be created through lot's of different types of media and art: sculptures, photography or film and also written portraits exist \cite{10.1093/aesthj/ayv018}. All these have in common that they want to capture essential features of the portrayed person like their appearance, personality, mood and sometimes also give an insight into the person's life story.\\
A portrait also always contains the artist's intent: this might be the truthful depiction of a person's life in a biography movie, emphasizing a person's beauty and wealth in classic historic paintings or exaggerating visual features in a caricature for entertainment. The missing intent is also why a quick photo snapshot usually is not considered a portrait.\\
\ \\
For this paper we are aiming to create joyful abstract portrait sketches from images. These sketches should be drawn on paper by a drawing robot and also have a unique visual style.\\
Drawing robots have become essential tools for the digital art community, especially in the genre of generative art. Examples of well-known art projects and experiments have been created by Jon McCormack et al. \cite{mccormack2017niche}, who built a swarm of small driving and drawing robots called ``DrawBots''. They utilize evolutionary simulation to give each robot it's own aesthetic preference. Another artist is Sougwen Chung \cite{Chung2022}, who uses large industry robot arms to co-creatively paint large abstract paintings. Finally there is also ``Sketchy'' by Jarkman \cite{Sketchy}, which is a small Arduino-based drawing robot that can take pictures and draw very rough facial sketches using edge detection.\\ 
\ \\
To give our portraits a unique aesthetic style, we aim to include generative patterns into the images as a form of shading. For this we draw single template sketches that act as reference for a neural net. This net is trained to learn the seen stroke shapes, recreate and also slightly alter them. These strokes are scattered in the darker areas of the image to create a unique shading style.\\
Freehand sketch representation learning, recognition and generation is a large and diverse research field \cite{xu2022deep}, where we mainly focus on one-shot learning and derivate generation. A lot of different forms of data representation have been presented over the last years, though as we aim to use a drawing robot as output medium, research that mainly focuses on vector representation is of interest here. Sketch-RNN \cite{ha2017neural} for example uses a recursive neural net and interprets a sketch drawing as sequence of turtle-graphic-like moves drawn by a virtual pen. A more modern approach is our prior work StrokeCoder \cite{wieluch2020strokecoder}, which uses a transformer to learn freehand sketch data and is also able to create derivates, though only in a restricted context.\\
So for this paper we utilize graph neural nets in combination with graph convolution \cite{kipf2017semisupervised}, which allows us to have a single vector representation of a full stroke and also benefit from the possibility of latent space exploration to generate stroke derivates.

\section{Generating Sketch Portraits: System Overview and Constrains}
\begin{figure}%
	\centering
	\subfloat[\centering Used Devices]{{\includegraphics[width=5.8cm]{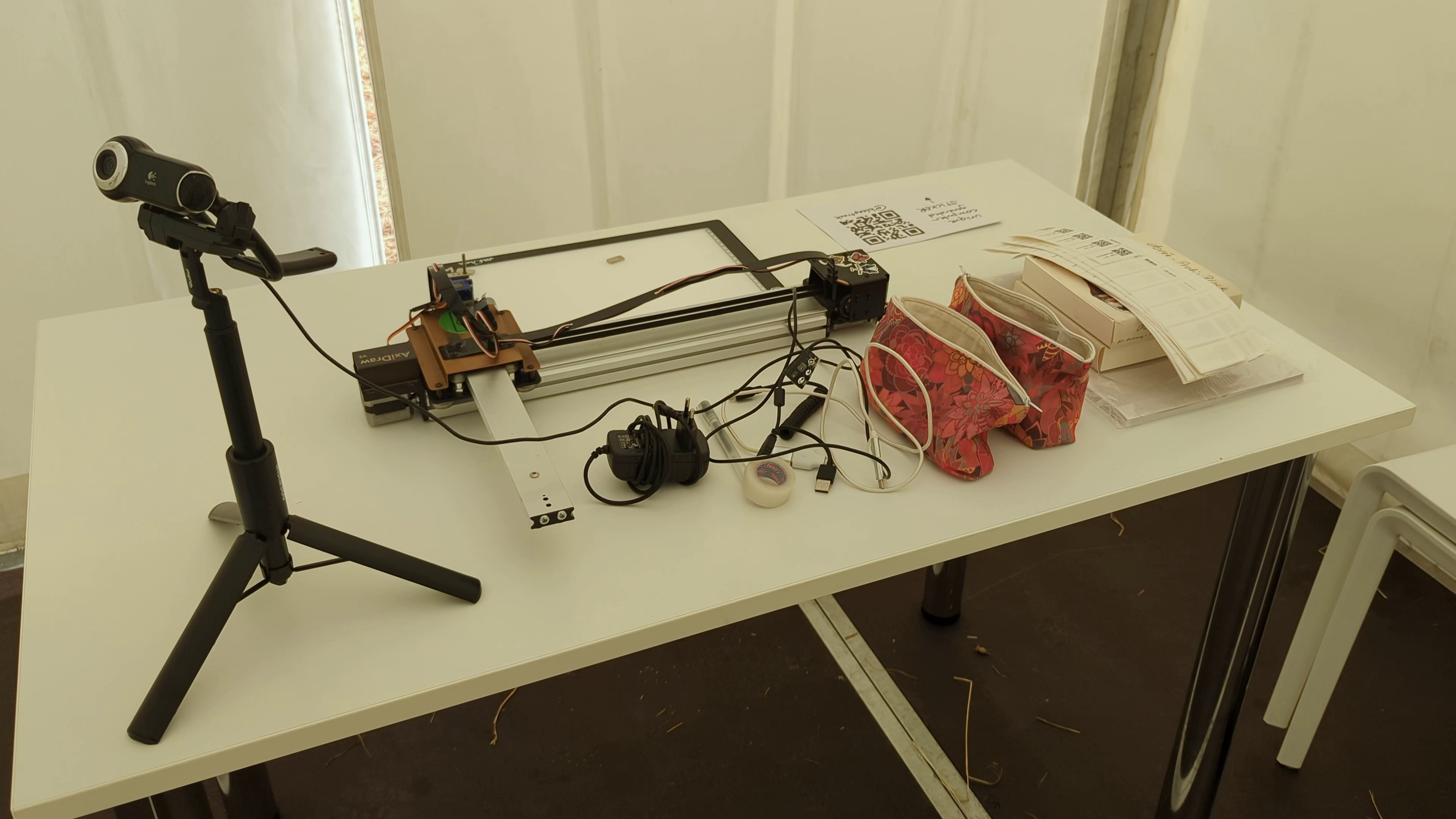} }}%
	\quad
	\subfloat[\centering Drawing in Process]{{\includegraphics[width=5.8cm]{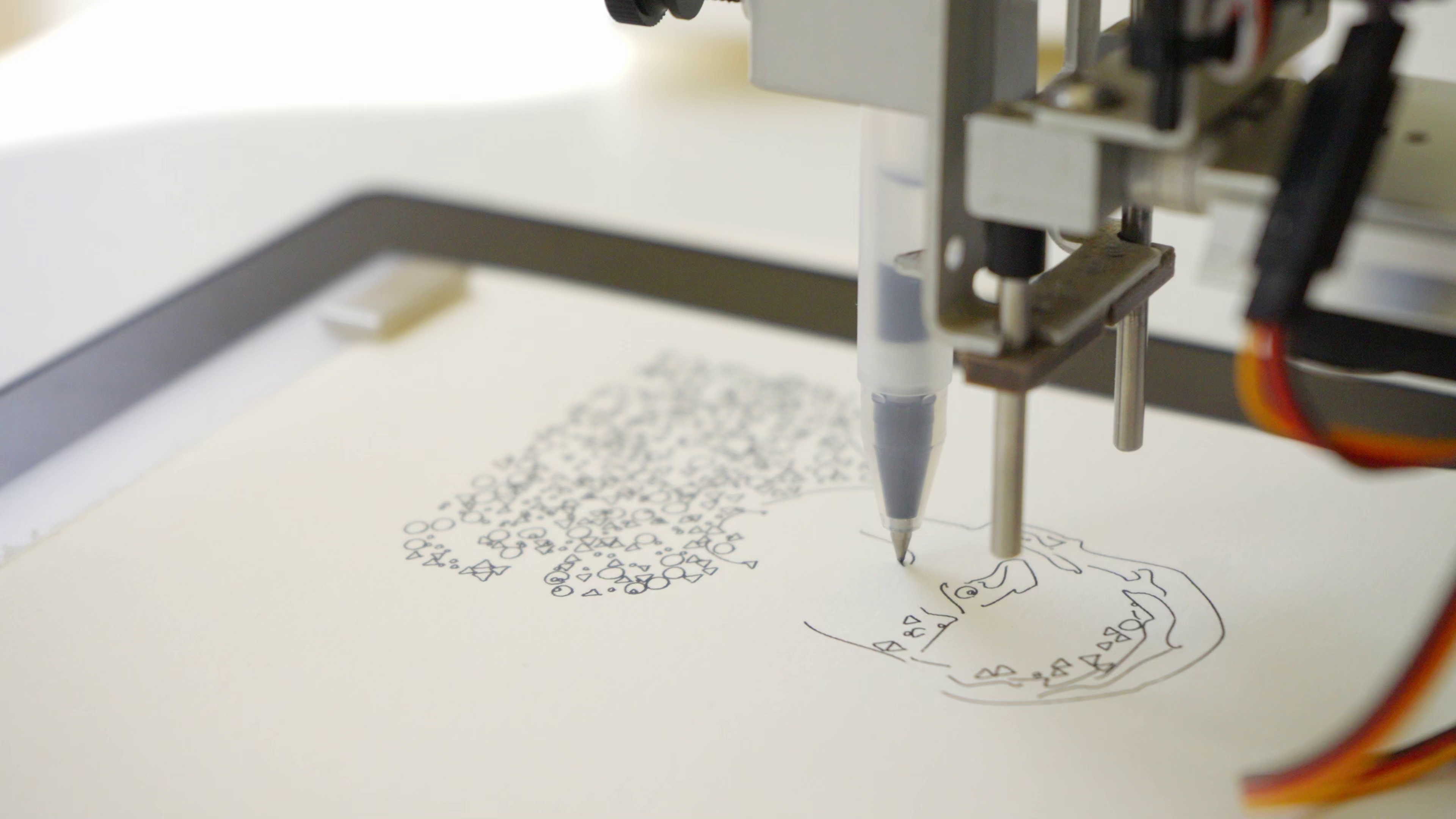} }}%
	\caption{``PatternPortrait'' setup: pictures are taken with a regular webcam in SD resolution. The drawings are created with an Axidraw V3 pen plotter.}%
	\label{fig:plotter}%
\end{figure}

``PatternPortrait'' was born from the idea to create a drawing machine that would be capable of taking a snapshot of a person's face and then draw a quick, simplified portrait from that, injecting a unique style by shading with different patterns.\\
Our setup for this installation consists of a webcam mounted on a tripod to take pictures, an Axidraw V3 drawing robot in combination with a magnetic board to hold the paper in place. Both devices are connected to a notebook running the application. Images of this setup can be seen in figure \ref{fig:plotter}.
\ \\
Derived from the previously stated idea, we identified the following constrains for creating such a drawing bot installation:
\begin{enumerate}
	\item The project's input is a regular pixel image but the output it drawn with a drawing robot (aka a pen plotter), therefore a transformation from one image type to the other is required.
	\item The overall process of taking the picture, calculating the output image and the drawing process itself should not take longer than approximately 10 minutes to prevent boredom and frustration for the viewer.
	\item The system should be capable of learning line styles from single template sketch drawings and arranging them into a simple pattern as shading. We aim for single template images, as we ourselves can not provide a dataset containing thousands of sketch drawings created by us. Also we promote the idea that AI systems should be transparent and ethical in their data usage, so we would like to construct systems that can work only on a single artist's data.
\end{enumerate}
These three constrains will act as guidelines in the following sketch portrait generation process.
\section{From Picture to Lines}
\label{pictoline}
The system's first task is to create lines from the initial input image. There are several approaches of creating artistic line drawings from images, for example through a reaction-diffusion \cite{pearson1993complex} simulation, where parameters are changed depending on the pixel lightness or by calculating points of a weighted Voronoi net \cite{secord2002weighted}, which are then connected to a single line via approximating the Traveling Salesman Problem. Both examples can be seen as a form of dithering, as these approaches aim to also represent the original image's brightness distribution. We instead aim for an algorithm that only translates the main visible features and shapes, like people's facial or body features, into lines and so more closely resembles a human approach of drawing a face.\\
\ \\
To achieve this translation from picture to lines, we first use canny edge detection \cite{canny1986computational} to retrieve a binary image that contains highlighted pixels belonging to a detected edge. In a next step, we need to identify pixels that visually belong to the same edge or line. This task is not trivial and several complex approaches \cite{vecto}\cite{vecto2} to solve this vectorization problem have been presented over the last years. Though for our artistic goal, we are not in need for a perfect exact result and therefore are using a very simple, greedy approach:\\
Every pixel is interpreted as a point $p \in P$ (with $P$ being all found pixel points), located at the pixel's $x$ and $y$ coordinate. A random $p$ is sampled from $P$ and it's closest neighbor point $p_{nbr} \in P$ is determined. If the distance between both points is smaller than a threshold $||p_{nbr} - p|| < d$, both points are connected to a path and removed from $P$.\\
Now this process is iteratively continued from both path ends: the next closest point $p_{nbr} \in P$ is detected and compared to the current path end point $||p_{nbr} - p{end}|| < d$. If the closest point to the path end is close enough, it is added to the path and so becomes the new end point and is also removed from $P$. This search is repeated until no point exists, that can ne added to the path or $P$ is empty. If this is true for both path ends, the path is considered finished and a new random point $p$ is sampled from the remaining points in $P$ until all points have been visited.\\
The so generated vector paths contain lot's of unnecessary points which even make the path look jittery on close inspection. For this reason all extracted paths are simplified \cite{schneider1990algorithm} via line fitting in a last step. 
\begin{figure}%
	\centering
	\subfloat[\centering Original Picture]{{\includegraphics[width=5.8cm]{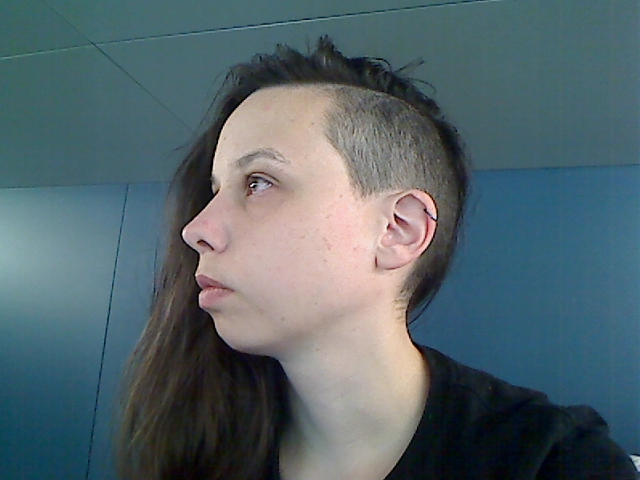} }}%
	\quad
	\subfloat[\centering Canny Edge Detection]{{\includegraphics[width=5.8cm]{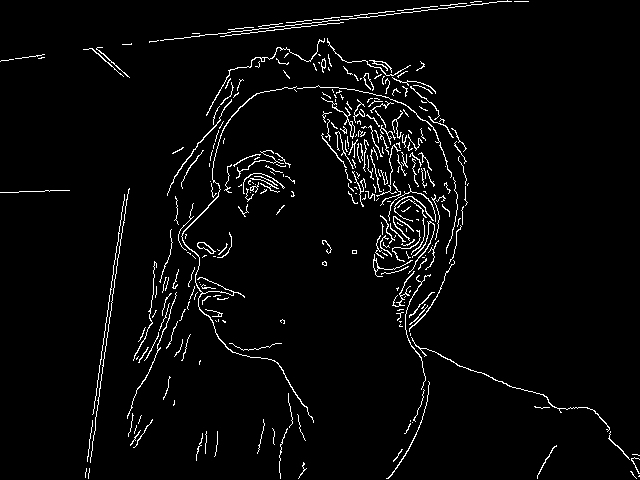} }}%
	\quad
	\subfloat[\centering Vectorization Result]{{\includegraphics[width=5.8cm]{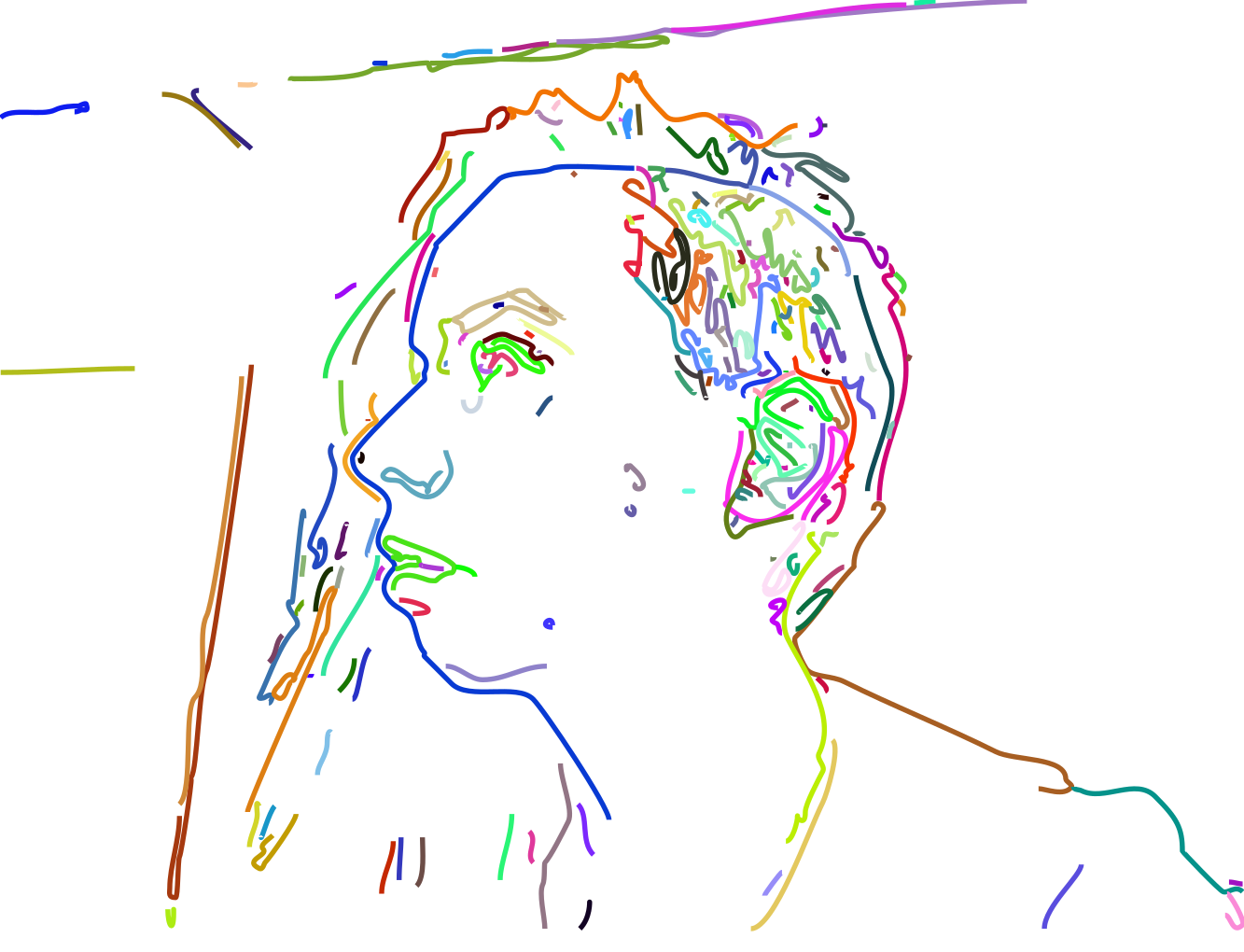} }}%
	\caption{Vectorization process: on the original picture a canny edge detection is applied that is then vectorized by the algorithm described in section \ref{pictoline}. Each vectorized line is assigned a random color for better differentiation.}%
	\label{fig:vector}%
\end{figure} 
\section{Generating Abstract Shading via Graph Neural Nets}
\label{graphstruct}
\begin{figure}%
	\centering
	\includegraphics[width=9cm]{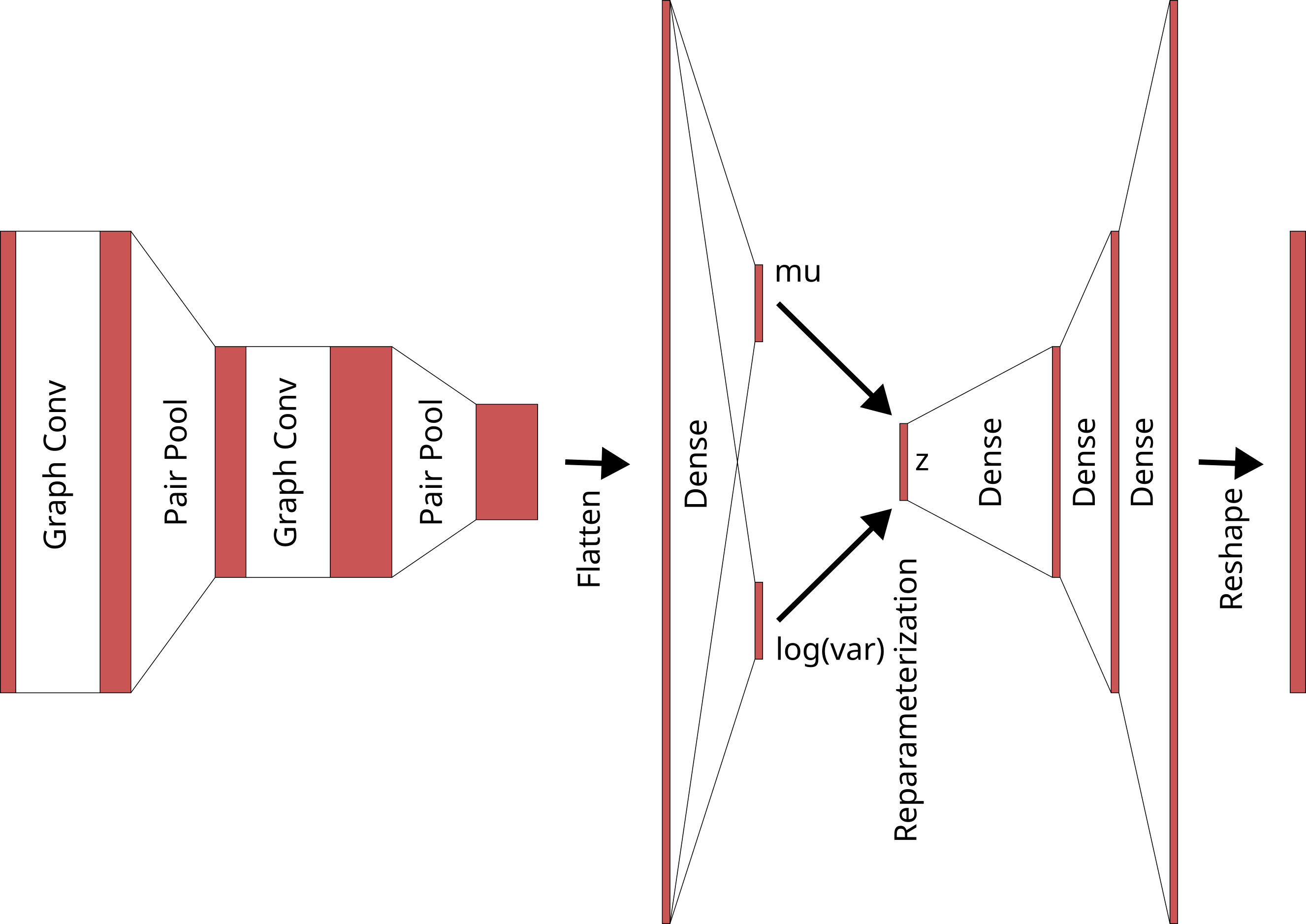} 
	\caption{variational Graph autoencoder Overview}%
	\label{fig:nnet}%
\end{figure}

After retrieving the vectorized canny edge detection lines, the only step missing is a form of shading. Here the goal is to search for dark areas in the original image and fill these areas with shapes generated from a single reference drawing.\\
\ \\
To always include the full stroke in context, we decided to interpret each stroke as an undirected graph $G(V,E)$ which consists of $n$ vertices $v$ with $v \in V$ that are equally distributed along the stroke line. All vertices $v_1 ... v_n$ are ordered in stroke direction, meaning that $v_1$ represents the stroke start and $v_n$ represents the stroke end. Also each vertex $v$ has positional attributes $<\Delta x,\Delta y>$ that contain the positional delta to it's predecessor. Vertex $v_1$ acts as reference point and so receives the positional attributes $<0,0>$.\\
The graph also consists of $2(n-1)$ edges $e \in E$ where each vertex $v_i$ is connected to the reference point $v_1$ and also to it's neighboring vertices $v_{i-1}$ and $v_{i+1}$ on the stroke line, if they exist. This graph structure allows for local neighborhood knowledge in each vertex $v$ but also for global knowledge collected in vertex $v_1$ when graph convolution is used. A visual representation of the graph structure can be seen in figure \ref{fig:graph} below:\\
\begin{figure}%
	\centering
	\includegraphics[width=7cm]{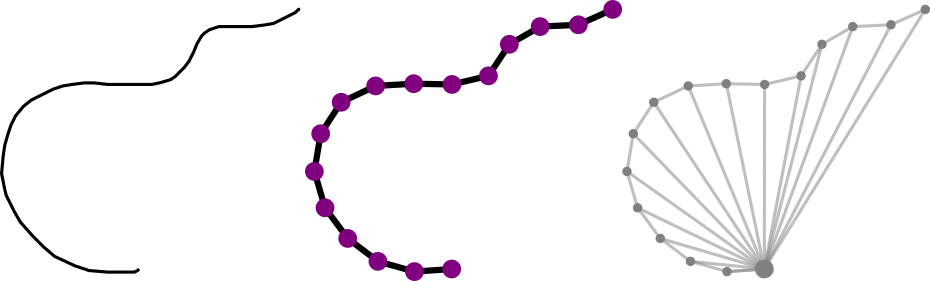} 
	\caption{Graph representation of a single stroke, with the original stroke on the left. Then partitioned into $n$ points of equal distance on the stroke and finally converted to a graph structure as described in section \ref{graphstruct}.}%
	\label{fig:graph}%
\end{figure}
\ \\
The graph representation is then trained with a Graph Encoder, that shows similar structures to a variational autoencoder (VAE) \cite{kingma2013auto}. The graph matrix of size $n \times 2$ (2 positional attributes $\Delta x$ and $\Delta y$) is processed through two blocks of a graph convolution layer and a pooling layer. The graph convolution layer used is the same as described by Kipf et al. \cite{kipf2017semisupervised}, which uses addition as the message passing aggregator. The pooling layer applies a pairwise one-dimensional max-pooling along the ordered vertices, which halves the matrix dimension. The result is then flattened and transformed into two vectors $\mu$ and $\sigma$ via a fully connected layer. With help of the reparametrization trick by Kingma et al. \cite{kingma2013auto}, the latent vector $z$ can be obtained from the distribution described by $\mu$ and $\sigma^2$.\\
The decoding process of $z$ is done via three fully connected layers and a final reshaping to receive the same dimensions on the output matrix as for the input.\\
The complete neural net structure can be found in figure \ref{fig:nnet}:\\
\ \\
In a final step, the color space of the original image is reduced to only very few colors by choosing the most dominant colors and matching each pixel to this color palette. To achieve this, we use modified median cut quantization as described by Frederick at al \cite{fredrick1992quantization}. Next, all pixels containing the darkest quantized color are chosen as possible area to fill with a pattern. In a final step, generated strokes from the trained autoencoder are placed on these pixels with the constraint that they must not touch any other already placed line. Utilizing the possibility to also generate slightly altered strokes helps the image to obtain a hand-drawn look. This process creates a random pattern that acts as shading in the portrait sketch, though not taking the stroke line placement of the original template sketch into account. We see many ways of possible improvements here, as described in future work.
	
\section{Results and Discussion}
\begin{figure}%
	\centering
	\includegraphics[width=12cm]{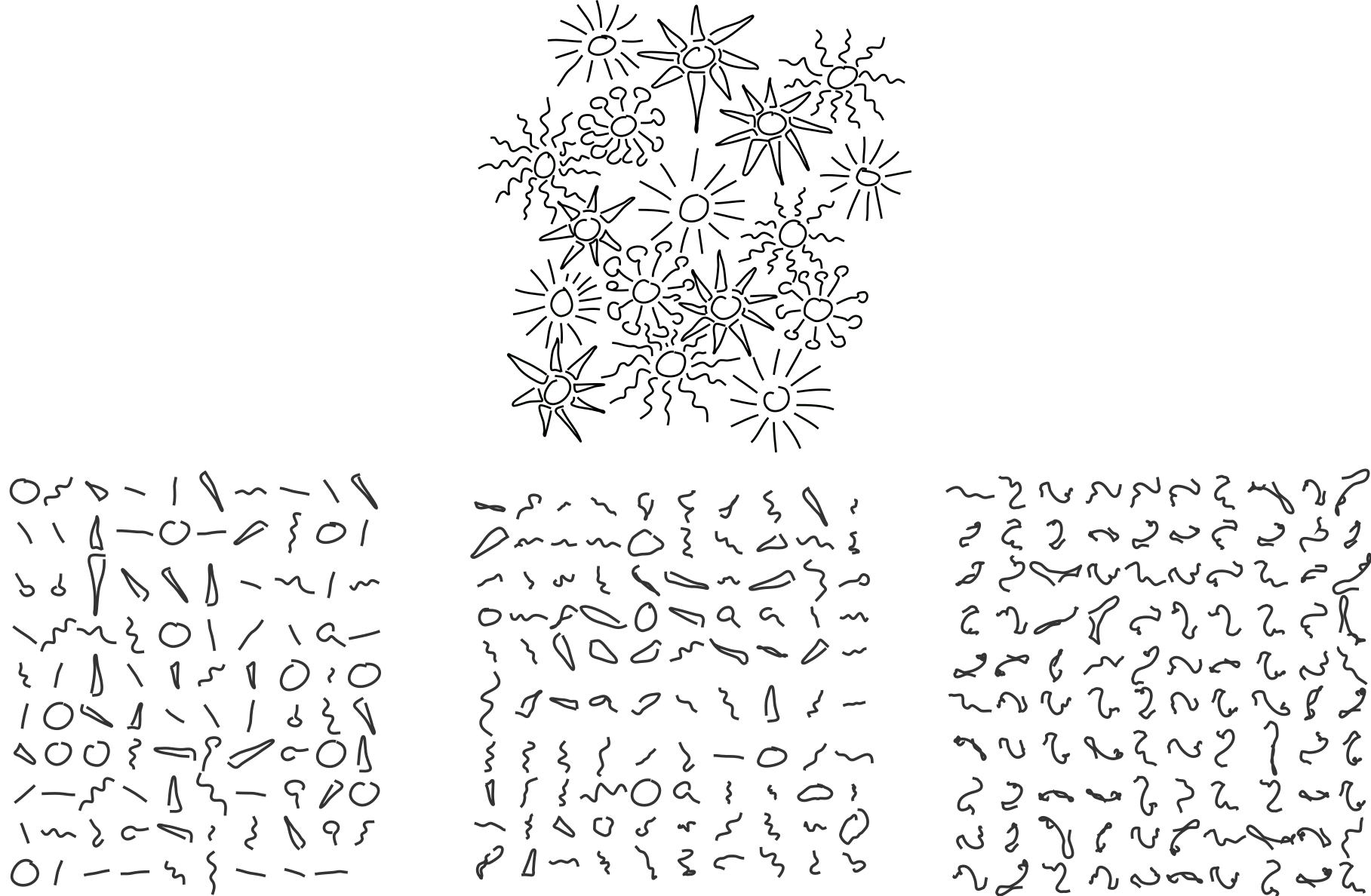} 
	\caption{Latent space observations: a model is trained with all strokes in the sample sketch shown at the top. A first (left) set is generated by observing the decoded original strokes. The second set (middle) is produced by adding noise to the original strokes latent vectors. And finally a last set (right) is created by sampling random latent vectors.}%
	\label{fig:latspace}%
\end{figure}

\begin{figure}[h]%
	\centering
	\subfloat{{\includegraphics[width=5cm]{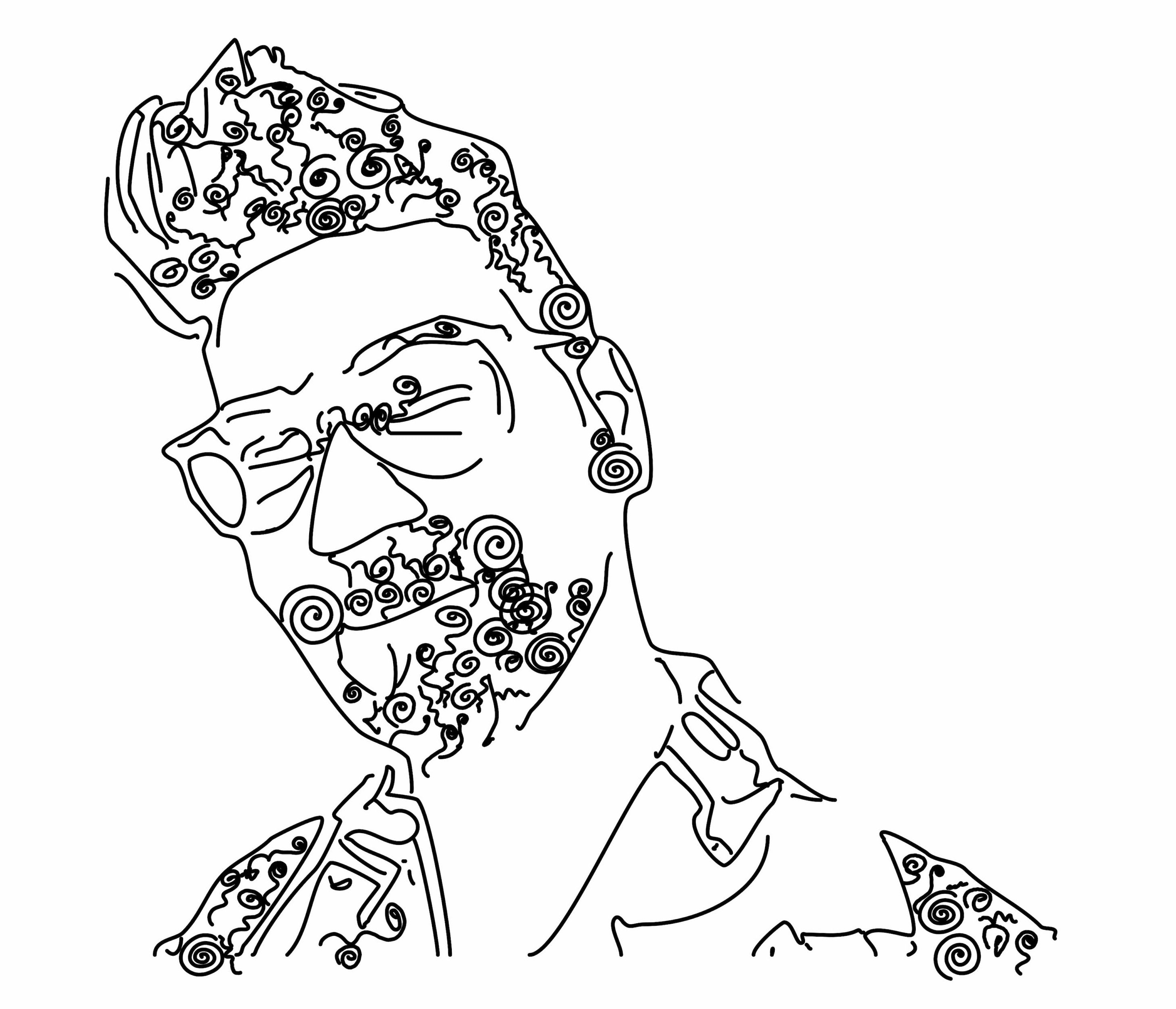} }}%
	\qquad
	\subfloat{{\includegraphics[width=5cm]{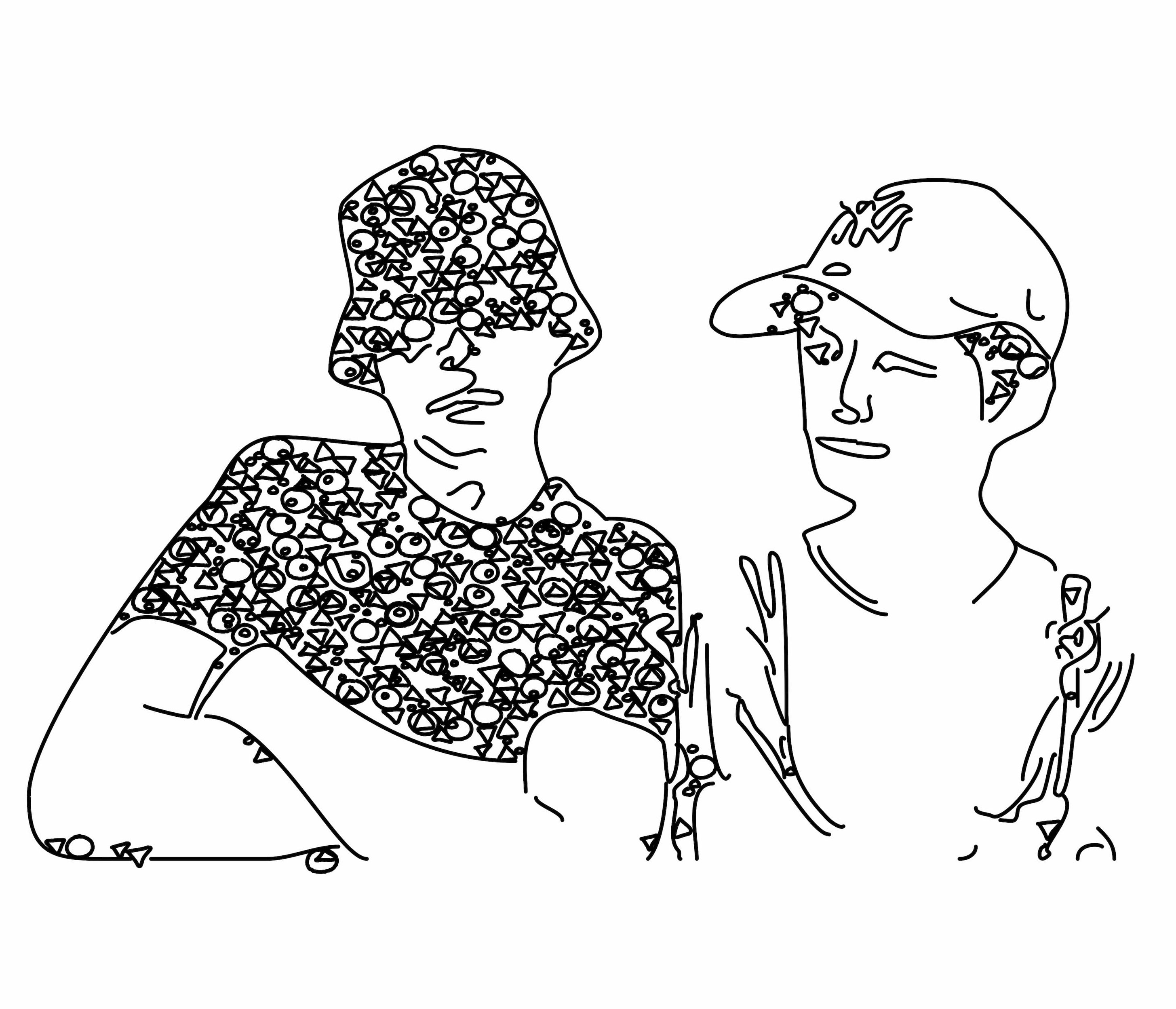} }}%
	\qquad
	\subfloat{{\includegraphics[width=5cm]{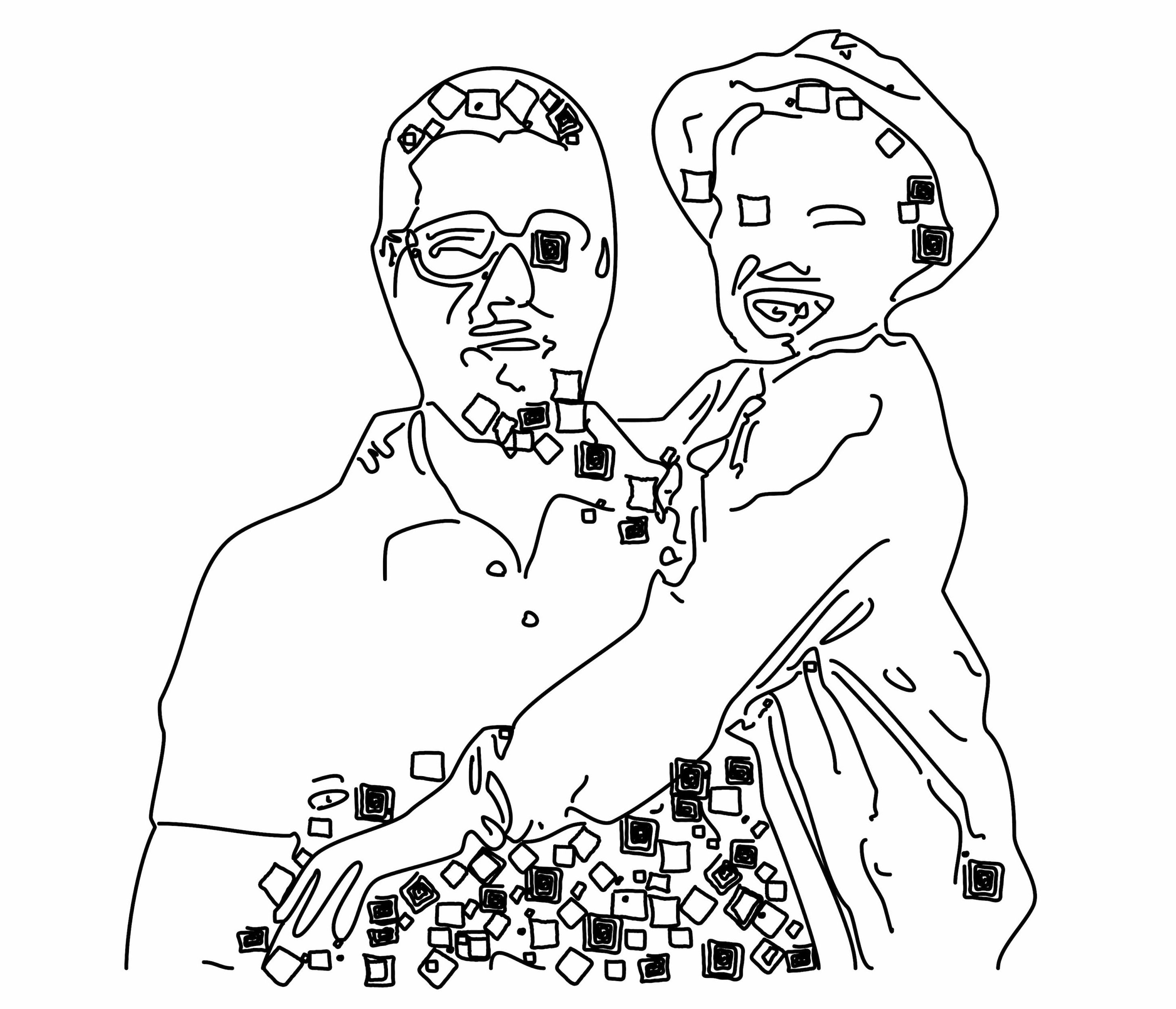} }}%
	\qquad
	\subfloat{{\includegraphics[width=5cm]{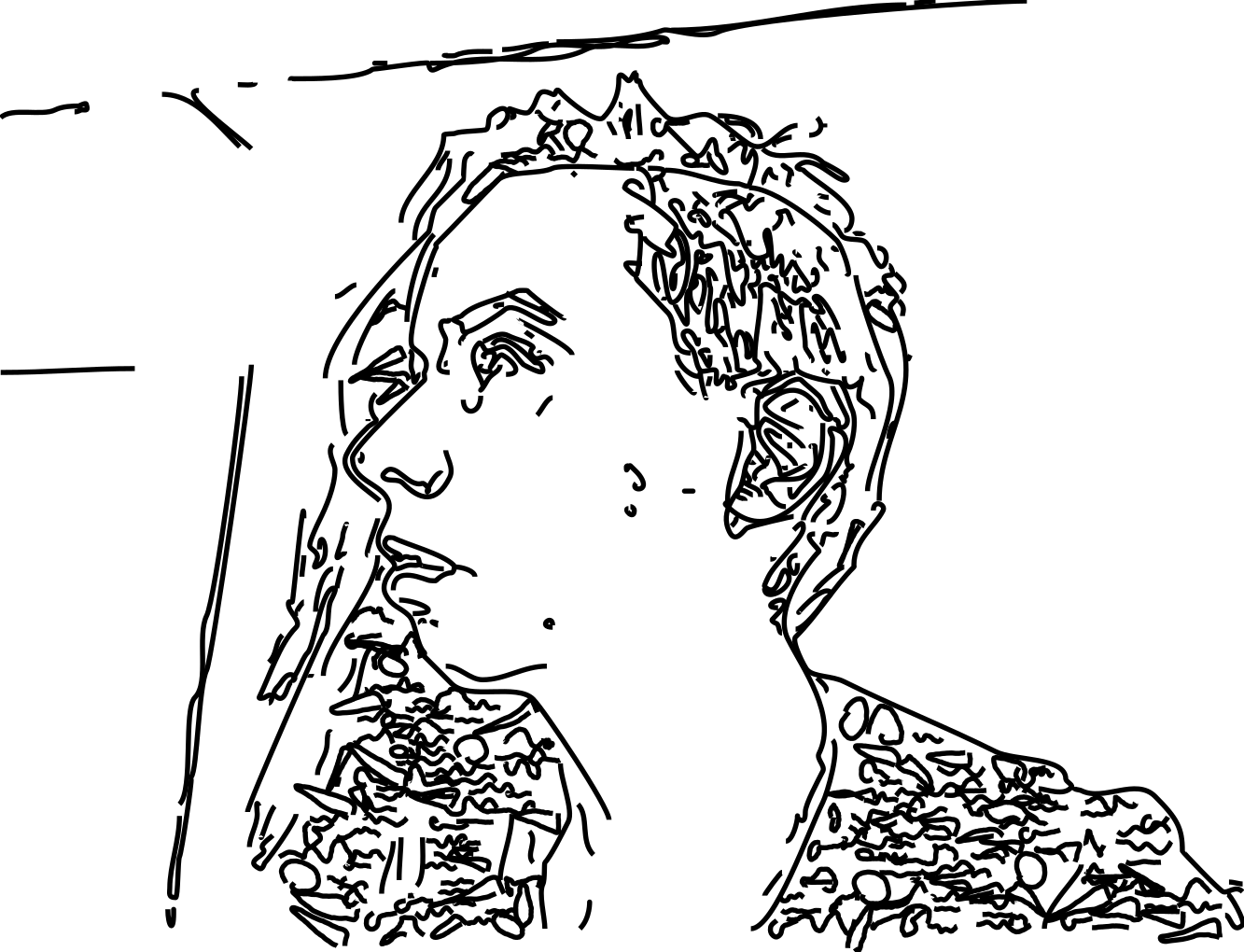} }}%
	\caption{Close up of four resulting ``PatternPortrait'' images. The bottom right image is the result from the process shown in Figure \ref{fig:vector}.}%
	\label{fig:closeup}%
\end{figure}

\begin{figure}%
	\centering
	\includegraphics[width=13cm]{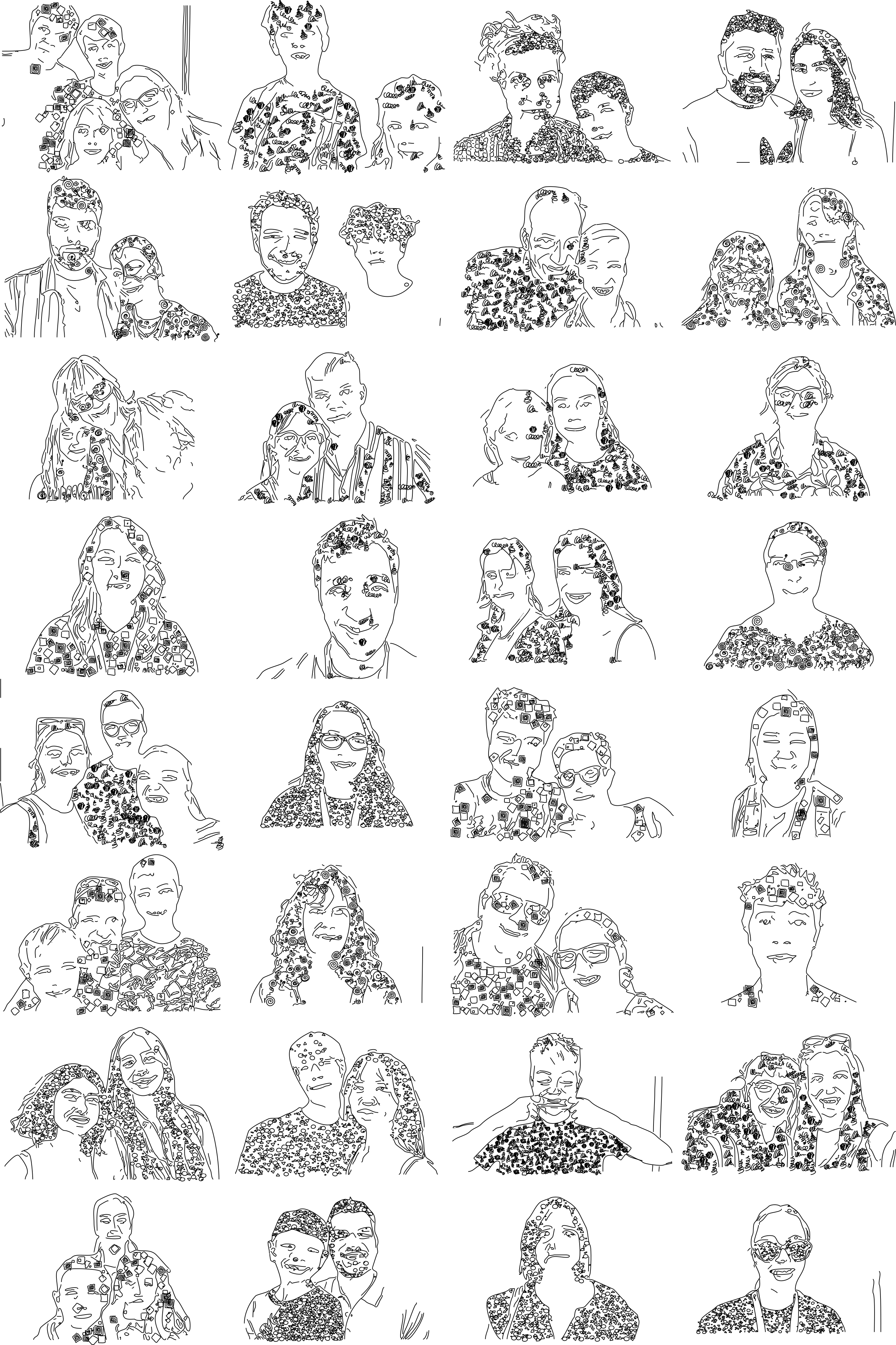} 
	\caption{Subset of ``PatternPortraits'' taken at a festival on a bright and sunny day with white plain background.}%
	\label{fig:grid}%
\end{figure}

With the use of variational autoencoders, the latent space is forced to represent a gaussian distribution whereas regular autoencoders create a manifold with unknown shape. This regularization comes with the benefit that latent space vectors can be sampled randomly and still represents to the learned model.\\
Figure \ref{fig:latspace} shows results from the trained vector line model: all lines in the top sample sketch have been used as training data. Their representation is learned well, as can be observed in the left line set: here the original lines are retrieved from the latent space and resemble the original shapes very well. It is now possible to add noise to the latent vectors to receive slight alterations of the original strokes. This works well, but also deformities can be observed (center set), especially visible on pointy triangles and circles. As mentioned above, it is possible to sample random latent vectors (right set). Here, most of the strokes are some mixed form of the original lines and do visually not resemble the original strokes well.\\
We decided to add slight noise to our latent vectors to benefit from the novel strokes but still keeping the original style recognizable.\\
\\
Final results can be seen in figure \ref{fig:grid}, which are a subset of all 280 PatternPortraits drawn so far. Figure \ref{fig:closeup} shows four additional close ups, containing the final image retrieved from the picture shown in figure \ref{fig:vector}.\\
\\
\begin{figure}%
	\centering
	\subfloat[\centering Poppy Seed Capsules]{{\includegraphics[width=3.7cm]{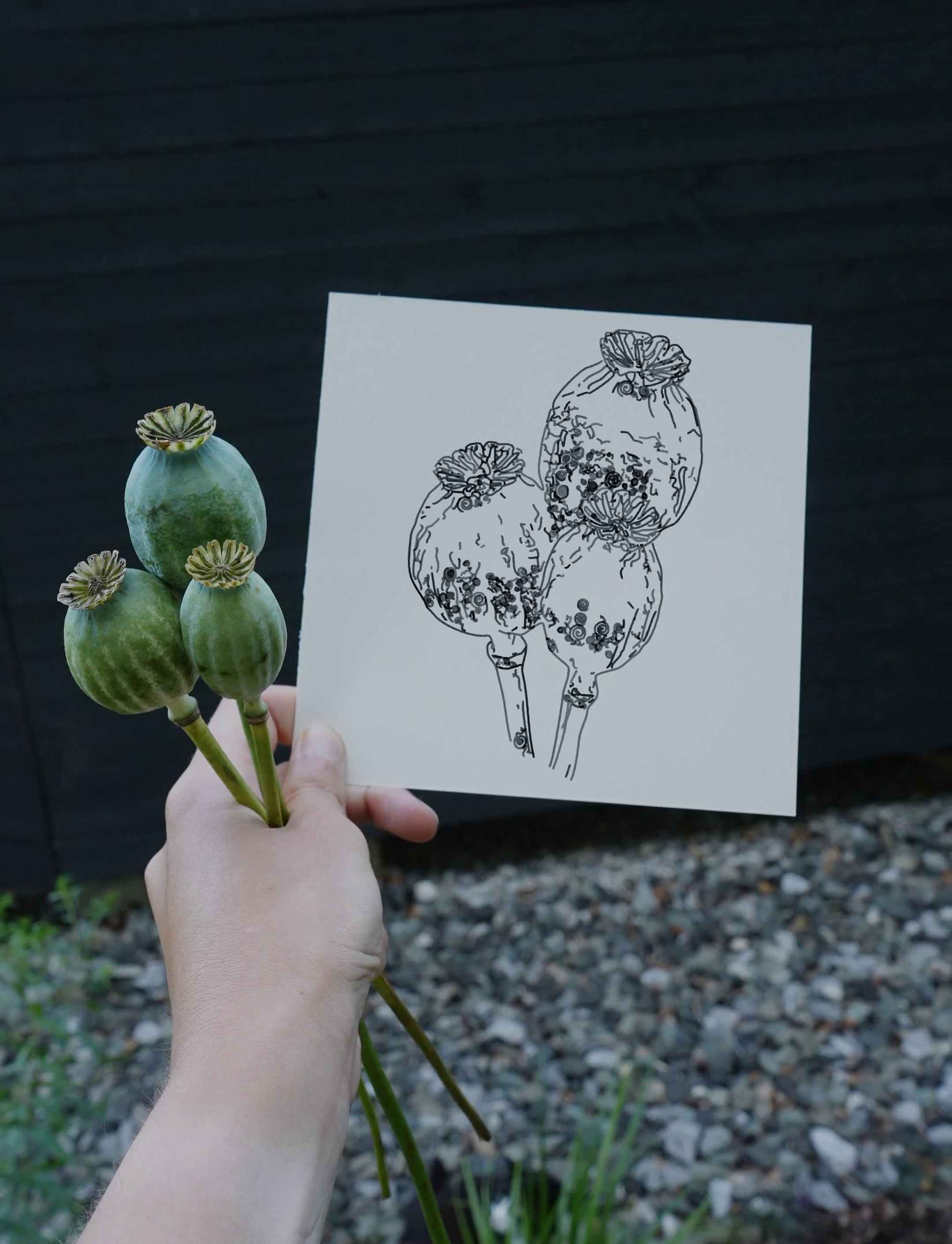} }}%
	\quad
	\subfloat[\centering Snail]{{\includegraphics[width=3.7cm]{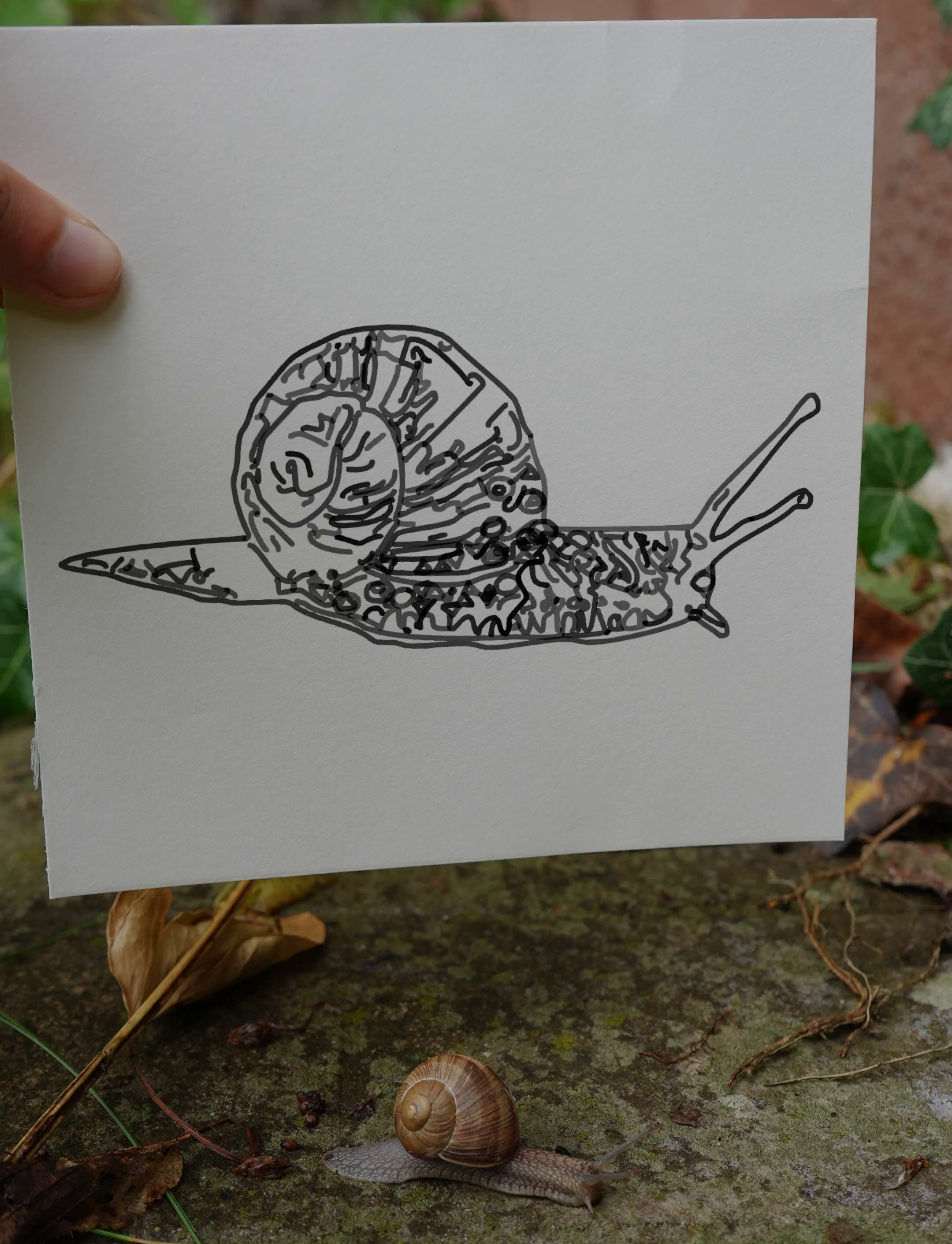} }}%
	\quad
	\subfloat[\centering Succulent Plant]{{\includegraphics[width=3.7cm]{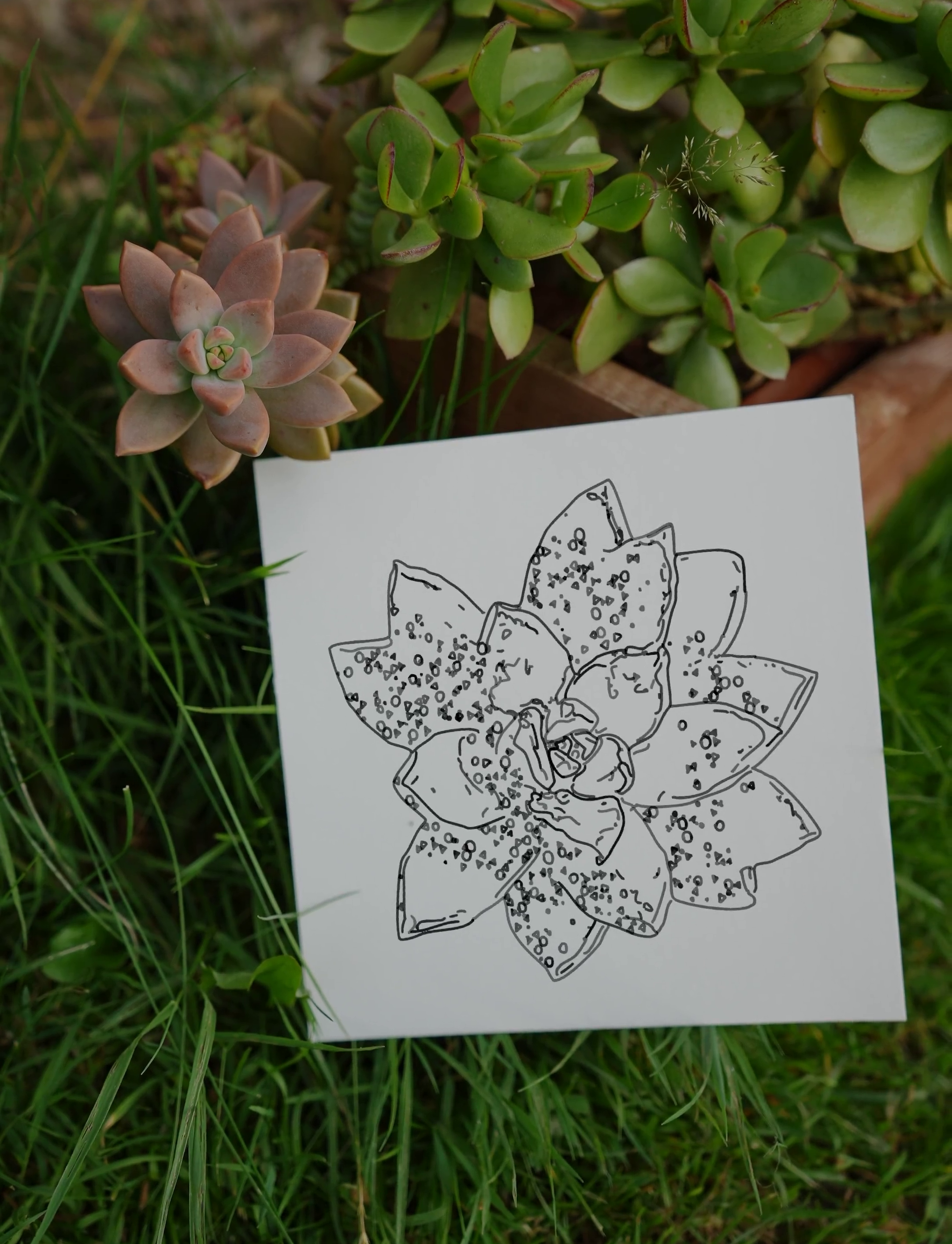} }}%
	\caption{``PatternPortrait'' experiments with animals and plants as picture subjects.}%
	\label{fig:vids}%
\end{figure}
\ \\
Besides using humans as depiction subjects, some further experiments were made with animals and inanimate objects. Though with an additional step of removing the subject's background manually, as the pictures were taken in visually very noisy environments. Exemplary results can be seen in \ref{fig:vids}.
\\
\ \\
While creating about 280 ``PatternPortraits'', we could observe recurring reactions from visitors that we summarize as qualitative results in the following section:\\
\begin{itemize}
	\item Either while plotting an image or when seeing a finished plot of another person: people were quick to identify and match the depicted face, often verbally discussing who is currently drawn or stating how well they think the person was captured in the final image.
	\item The pen plotter drew the final images by choosing the fastest way through all lines. This drawing order was often verbally described as non-human and even confusing.
	\item Most visitors very much enjoyed the abstract image style with all it's quirks and imperfections. Though also some visitors mentioned they did not feel represented well enough, especially when some of their facial featured weren't captured well by the edge detection or a pattern stroke was placed in a way where it interfered with their facial features. For example a large stroke is placed in the eyebrow area resulting in a visually heavily enlarged eyebrow.
	\item People enjoyed posing and experimenting in front of the camera and were overall curious to see the results. It occurred multiple times that a person created a ``PatternPortrait'' to then later come back with their friend group to create more. We interpret behavior as very positive and joyful attitude towards the project.
\end{itemize}
Besides visitor's reactions, we also would like to note some technical details:
\begin{itemize}
	\item The canny edge detection can be run in real time, which es very helpful to give a preview to the visitors and overall for proper framing of the picture.
	\item The used webcam with low SD resolution (640 by 480 pixels) is enough to create ``PatternPortraits''. So no additional expensive camera equipment is needed.
	\item The canny edge detection kernel size and also the pattern stroke size and number are parameters that can be used to tune the overall process time from taking the image to finished pen plot. We tuned the process to have an overall process time between 4 and 10 minutes, which we found a suitable waiting time for visitors while maintaining enough detail in the image.   
	\item The vectorization process is rather slow, especially when visitors wear high-contrast clothing, which creates lots of extra lines to calculate. This could be optimized by either using a faster algorithm or even train a neural net to perform the task.
	\item The perceived gray value from the resulting patterns does not match the original image's perceived darkness. Also the pattern's gray value can shift dramatically depending on the strokes used and their currently random placement. Ways to optimize this behavior is discussed in the next section containing ideas for future work.
\end{itemize}

\section{Conclusion and Future Work}
In this paper we presented a process to create abstract portrait drawings from images that playfully utilize single freehand sketches as reference to create shading patterns. We showed a strategy to extract facial and body features and transform then into vector lines. Additionally we presented a graph neural net architecture that is capable of learning sketch stroke representations in their vector represented form and can generate variations of those strokes. Combining these two approaches results in interesting and joyful abstract drawings that were well received by about 280 participants.\\
While already receiving very positive feedback on the generated portraits, this project only resembles the first step into the creative usage of generative pattern models.\\
A possible next step is to not only learn the single line representations but also the overall arrangement of lines in the template sketch. This would allow to experiment with different (co-)creative tasks like automatic extension of patterns, outer shape-dependent pattern filling or the creating of co-creative drawing tools \cite{wieluch2021co}.\\
Continuing in the line of transferring pictures to abstract line drawings, it would be important to better control the gray value. This would not only heavily improve the overall image quality but would also be an interesting parameter to use in the above mentioned (co-)creative tasks.\\
A further improvement directly regarding ``PatternPortrait'' would be to include image segmentation to remove dark areas from the background. This would ensure that the depicted person is always the focus point and no unnecessary lines or patterns are created. Additionally, the face region could be excluded from the area of possible pattern placement to benefit the facial feature depiction and also prevent unnecessary pattern placement in case of a darker skin.\\
While in this project it would also have been possible to interpolate between sketch lines to retrieve derivative stroke lines, variational autoencoder and similar architectures encode more information than only the pure shape. Instead similar elements are clustered together. We hope to exploit this feature in future work for learning more complex patterns. Another interesting approach for future generative systems would be the implementation of diffusion models on vector-based sketch data \cite{wang2023sketchknitter}.
%
%
%
\newpage
\bibliographystyle{splncs04}
\bibliography{mybibliography}

\end{document}